\documentclass[graybox]{svmult}

\usepackage{graphicx}

\usepackage{amsmath}
\usepackage{amssymb}
\usepackage{color}

\usepackage[latin1]{inputenc}
\usepackage{amsfonts}
\usepackage{tikz} 
\usetikzlibrary{matrix}
\usepackage[english]{babel}
\usepackage[latin1]{inputenc}
\usepackage[T1]{fontenc}
\usepackage{ae}

\usepackage{url}

\allowdisplaybreaks[4]

\usepackage{color}
\usepackage{tikz}
\usetikzlibrary{matrix}
\usepackage{cite}           
\usepackage{mathptmx}       
\usepackage{helvet}         
\usepackage{courier}        
\usepackage{type1cm}        
\usepackage{makeidx}         
\usepackage{graphicx}        
\usepackage{multicol}        
\usepackage[bottom]{footmisc}

\usepackage[fonts]{mpostinl}
\usepackage{graphbox}

\newtheorem{thm}{Theorem}
\newtheorem{defn}{Definition}

\newtheorem{conj}{Conjecture}
\newtheorem{prop}{Proposition}

\DeclareMathOperator{\AcycLetter}{\text{\textbf{A}}}
\DeclareMathOperator{\BcycLetter}{\text{\textbf{B}}}
\DeclareMathOperator{\DcycLetter}{\text{\textbf{D}}}

\DeclareMathOperator{\smallBcycLetter}{\text{\textbf{b}}}
\DeclareMathOperator{\smallDcycLetter}{\text{\textbf{d}}}
\newcommand{\Acyc}[1]{\AcycLetter\mkern-4.2mu\Big[\mpostuse[width=0.5cm,align=c]{#1}\Big]}
\newcommand{\Bcyc}[1]{\BcycLetter\mkern-4.2mu\Big[\mpostuse[width=0.5cm,align=c]{#1}\Big]}
\newcommand{\Dcyc}[1]{\DcycLetter\mkern-4.2mu\Big[\mpostuse[width=0.5cm,align=c]{#1}\Big]}

\newcommand{\smallBcyc}[1]{\smallBcycLetter\mkern-4.2mu\Big[\mpostuse[width=0.5cm,align=c]{#1}\Big]}
\newcommand{\smallDcyc}[1]{\smallDcycLetter\mkern-4.2mu\Big[\mpostuse[width=0.5cm,align=c]{#1}\Big]}

\usetikzlibrary{patterns,matrix,through,arrows,decorations.pathreplacing,decorations.markings,snakes,shadows,shapes.geometric,positioning,calc,backgrounds,shapes.arrows,fit,backgrounds}

\tikzset{anchorbase/.style={baseline={([yshift=-0.5ex]current bounding box.center)}},
int/.style={thick},
  cross line/.style={preaction={draw=white,line width=6pt,-}},
  wall/.style={thin,double,blue},
  middlearrow/.style={postaction=decorate,decoration={markings,mark=at
    position .55 with {\arrow{stealth};}}},
  middlearrowrev/.style={postaction=decorate,decoration={markings,mark=at
    position .55 with {\arrowreversed{stealth};}}},
  ev/.style={shape=rectangle, draw},
  every path/.style={line width=1pt}
}  
\def \myweightstyle {}
\def \myweightx {0.09}
\def \myweighty {0.2}
\newcommand{\mydrawdown}[1]%
  {\draw [\myweightstyle] #1++(0,-\myweighty) -- ++(\myweightx,\myweighty);%
    \draw [\myweightstyle] #1++(0,-\myweighty) -- ++(-\myweightx,\myweighty);}
\newcommand{\mydrawup}[1]%
  {\draw [\myweightstyle] #1 -- ++(-\myweightx,-\myweighty);%
    \draw [\myweightstyle] #1 -- ++(\myweightx,-\myweighty);}

\begin{mpostdef}
  xu:=0.8cm;
  ds:=3.2pt;
  ls:=0.6pt;
  pair vp[];
  path c;
  def pensize(expr s)=withpen pencircle scaled s enddef;
  def poly(expr k,offset)=
    for i=1 upto k:
    vp[i]:=0.5*xu*(dir ((i-1)*360/k+offset));
    endfor;
  enddef;
  def singleline(expr i,j)=
    draw vp[i]--vp[j] pensize(ls);
  enddef;
  def doubleline(expr i,j)=
    draw vp[i]{(vp[j]-vp[i]) rotated 25}..{(vp[j]-vp[i]) rotated -25}vp[j] pensize(ls);
    draw vp[i]{(vp[j]-vp[i]) rotated -25}..{(vp[j]-vp[i]) rotated 25}vp[j] pensize(ls);
  enddef;
  def tripleline(expr i,j)=
    draw vp[i]{(vp[j]-vp[i]) rotated 35}..{(vp[j]-vp[i]) rotated -35}vp[j] pensize(ls);
    draw vp[i]{(vp[j]-vp[i]) rotated 0}..{(vp[j]-vp[i]) rotated 0}vp[j] pensize(ls);
    draw vp[i]{(vp[j]-vp[i]) rotated -35}..{(vp[j]-vp[i]) rotated 35}vp[j] pensize(ls);
  enddef;
  def dottedline(expr i,j,num,ang)=
  c:= vp[i]{(vp[j]-vp[i]) rotated ang}..{(vp[j]-vp[i]) rotated (-1*ang)}vp[j];
    draw c pensize(ls);
    for vv=1 upto (num-1):
      draw point(vv/num) of c pensize(ds);
    endfor;
  enddef;
  def drawvert(expr i,j)=
    for vv=i upto j:
      draw vp[vv] pensize(ds);
    endfor;
  enddef;
  def drawcentralvert=
    draw (0xu,0xu) pensize(ds);
  enddef;
  def background =
    c:= (-1xu,-1xu)--(-1xu,1xu)--(1xu,1xu)--(1xu,-1xu)--cycle;
    c:=c scaled 0.52;
    fill c withcolor op*white;
  enddef;
  def framed(expr p)=fill bbox p withcolor white; draw bbox p pensize(0.8pt); draw p; enddef;
  def framedd(expr p)=fill bbox p withcolor 0.94white; draw bbox p pensize(1pt); draw p; enddef;
\end{mpostdef}

\begin{mpostfig}[label=myfig]
  poly(4,45);
  doubleline(1,2);
  tripleline(2,3);
  dottedline(3,4,3,20);
  for i=1 upto 4:
   draw vp[i] pensize(ds);
  endfor;
\end{mpostfig}

\begin{mpostfig}[label=empty]
  background;
\end{mpostfig}

\begin{mpostfig}[label=G2]
  background;poly(2,0);dottedline(1,2,0,35);dottedline(1,2,0,-35);drawvert(1,2);
\end{mpostfig}

\begin{mpostfig}[label=G3]
  background;poly(2,0);dottedline(1,2,0,-50);dottedline(1,2,0,0);;dottedline(1,2,0,50);drawvert(1,2);
\end{mpostfig}

\begin{mpostfig}[label=G111]
  background;poly(3,90); singleline(1,2);singleline(1,3);singleline(2,3); drawvert(1,3);
\end{mpostfig}

\begin{mpostfig}[label=G211]
  background; poly(3,90); singleline(1,2); singleline(1,3); doubleline(2,3); drawvert(1,3);
\end{mpostfig}

\begin{mpostfig}[label=GC211]
  background; poly(2,0);drawvert(1,2); dottedline(2,1,1,70); dottedline(2,1,1,-10); dottedline(2,1,2,-70); \end{mpostfig}

\begin{mpostfig}[label=G2111]
  background; poly(4,45); singleline(1,2); doubleline(2,3); singleline(3,4); singleline(4,1); drawvert(1,4);
\end{mpostfig}

\begin{mpostfig}[label=G321]
  background; poly(3,90); tripleline(1,2); doubleline(2,3); singleline(3,1); drawvert(1,3);
\end{mpostfig}

\begin{mpostfig}[label=G511]
  background; poly(3,90); 
  dottedline(2,3,0,-60);dottedline(2,3,0,-30);dottedline(2,3,0,0);dottedline(2,3,0,30);dottedline(2,3,0,60);
  singleline(1,2); singleline(1,3); drawvert(1,3);
\end{mpostfig}

\begin{mpostfig}[label=tetrahedral]
  background;poly(3,90); 
  vp[4]:=(0xu,0xu);
  singleline(1,2);singleline(1,3);singleline(2,3);drawvert(1,4);
  singleline(1,4);singleline(2,4);singleline(3,4); 
\end{mpostfig}

\begin{mpostfig}[label=test]
  background;
  poly(10,0);drawvert(1,10);
  doubleline(8,9);
  tripleline(9,10); 
  dottedline(1,6,4,-30);
\end{mpostfig}

\begin{mpostdef}
xu:=1cm;
psf:=0.85;
ahlength:=5pt;
string fftext;
qcdpos:=3;
nefpos:=1.25;
ostpos:=-0.65;
grpos:=-2.7;
\end{mpostdef}

\makeindex             

\begin{document}

\title*{Modular and holomorphic graph functions from superstring amplitudes}
\author{Federico Zerbini}
\institute{Federico Zerbini,\\  Institute de Physique Th\'eorique (IPHT), Orme des Merisiers batiment 774, Point courrier 136, F-91191 Gif-sur-Yvette Cedex, France,
\email{federico.zerbini@ipht.fr}
}
%
%
\maketitle

\begin{flushright}
  \verb!IPHT-t18/089!
\end{flushright}

\abstract{In these notes, based on a talk given at the conference \emph{Elliptic integrals, elliptic functions and modular forms in quantum field theory} held at DESY-Zeuthen in October 2017, we compare two classes of functions arising from genus-one superstring amplitudes: modular and holomorphic graph functions. We focus on their analytic properties, we recall the known asymptotic behaviour of modular graph functions and we refine the formula for the asymptotic behaviour of holomorphic graph functions. Moreover, we give new evidence of a conjecture appeared in~\cite{BSZ} which relates these two asymptotic expansions.}


%
%

\section{Introduction}
\label{sec:1}

\vspace{1mm}\noindent

The computation of the perturbative expansion of superstring scattering amplitudes constitutes an extremely fertile field of interaction between mathematicians and theoretical physicists. From a mathematician's viewpoint, this is partly due to the appealing and simple form of the Feynman-like integrals appearing in this expansion, but most remarkably to the fact that the mathematics unveiled is right at the boundary of our present knowledge: it constitutes both an astonishing concrete example of certain new abstract constructions from algebraic geometry, and at the same time it points towards new frontiers to be investigated. In this paper, besides presenting a detailed review of the state of art of number theoretical aspects of superstring amplitudes,
we will recall and refine recent results and conjectures presented in~\cite{BSZ} and~\cite{ZerbiniThesis}. We hope that this will be useful to understand the connection between genus-one superstring amplitudes and a new class of functions introduced by Brown in the context of his research on mixed modular motives~\cite{BrownNewClassI}. 

It was observed in~\cite{BSZ} that the real-analytic single-valued functions arising from genus-one closed-string amplitudes, called modular graph functions, seem to be related in a simple and intriguing way to a seemingly artificial combination of the holomorphic multi-valued functions arising from genus-one open-string amplitudes. This would extend the observation already made at genus zero that closed-string amplitudes seem to be a single-valued version of open-string amplitudes. After introducing some mathematical  background and recalling the genus-zero case, we will define modular and holomorphic graph functions, originating from genus-one closed and open-string amplitudes, respectively. Holomorphic graph functions are divided into A-cycle and B-cycle graph functions. The main theorems are the following:
\begin{thm}\label{Thm1}
The modular graph function associated to a graph $\Gamma$ with $l$ edges has the asymptotic expansion
\begin{equation*}
D_{\Gamma}(\tau)=\sum_{k=1-l}^l\,\sum_{m,n\geq 0}d_k^{(m,n)}(\Gamma)\,y^kq^m\overline{q}^n,
\end{equation*}
where $\tau\in\mathbb{H}$, $q=\exp(2\pi i\tau)$, $y=\pi\,Im(\tau)$ and the coefficients $d_k^{(m,n)}(\Gamma)$ are cyclotomic multiple zeta values.
\end{thm}
\begin{thm}\label{Thm2}
The B-cycle graph function associated to a graph $\Gamma$ with $l$ edges has the asymptotic expansion
\begin{equation*}
B_{\Gamma}(\tau)=\sum_{k=-l}^l\,\sum_{m\geq 0}b_k^{(m)}(\Gamma)\,T^kq^m,
\end{equation*}
where $T=\pi i\tau$ and the coefficients $b_k^{(m)}(\Gamma)$ are multiple zeta values.
\end{thm}
Theorem~\ref{Thm1} is a generalization of the main result of~\cite{Zerb15} and is already contained in the PhD thesis \cite{ZerbiniThesis}, while Theorem~\ref{Thm2} builds on previous results contained in~\cite{ZerbiniThesis} and~\cite{BSZ} but is ultimately new: its novelty consists in the explicit bound on the powers of~$T$ in terms of the number of edges~$l$ of the graph, which is obtained by exploiting an explicit formula for the open-string propagator.

In~\cite{Zerb15} it is conjectured that all $d_k^{(m,n)}(\Gamma)$ belong to a small subset of multiple zeta values called single-valued multiple zeta values. One of the main achievements of~\cite{BSZ} is to give great evidence of a (partly) stronger statement, which relates open-string amplitudes to closed-string amplitudes:

\begin{conj}
Let $sv:\zeta(\textbf{k})\rightarrow\zeta_{\rm sv}(\textbf{k})$. Then for any graph~$\Gamma$ with~$l$ edges and all $-l<k<l$ we have\footnote{The reason for the rational coefficient appearing in eqn.~(\ref{eq1}) is that we want to follow the notation adopted in the modular graph function literature. Setting $y=-2\pi\, Im(\tau)$, which mathematically would be a more natural choice, one would get a cleaner statement.}
\begin{equation}\label{eq1}
sv(b_k^{(0)}(\Gamma))=(-2)^{-k}d_k^{(0,0)(\Gamma)}.
\end{equation}
\end{conj}

We call this the (Laurent-polynomial) \emph{esv conjecture}, where the ``$e$'' stands for elliptic. This is not the only relationship observed in~\cite{BSZ} between holomorphic and modular graph functions; in particular, it seems that for graphs with up to six edges one can obtain modular graph function from holomorphic graph functions by applying a very simple set of ``\emph{esv} rules''. However, we prefer to be cautious and content ourselves to consider here and call conjecture only the Laurent-polynomial version.

In the last section of the paper we will recall the construction due to Brown of a single-valued analogue of certain holomorphic functions on the upper half plane $\mathbb{H}$ (iterated Eichler integrals of modular forms). We will argue that this construction should be related to our $esv$ conjecture, and mention various open questions and possible directions originating from this observation.

\section{Feynman periods}
\label{sec:2}

\vspace{1mm}\noindent

One of the origins of the interaction between physicists working on scattering amplitudes and mathematicians is the fact that Feynman integrals of quantum field theories are \emph{period functions} of the kinematic variables (e.g. masses, momenta). 

\emph{Periods} can be defined elementarily as absolutely convergent integrals of algebraic functions over domains given by polynomial inequalities with integer coefficients \cite{KontsZagier}. When the integrands depend algebraically on parameters, we call the integrals period functions, and for algebraic values of the parameters we get actual periods. Concretely, periods constitute a countable subring $\mathcal{P}$ of $\mathbb{C}$, which contains $\overline{\mathbb{Q}}$ as well as many (often conjecturally) transcendental numbers of geometric or number theoretical interest, such as $\pi=\iint_{x^2+y^2\leq 1}dxdy$ or special values of various kinds of L-functions. It is important to remark that the representation of a period as an integral is far from being unique: for instance, one also has $\pi=\int_{-1\leq x\leq 1}(1-x^2)^{-1/2}dx$. As a consequence of this, it is often very difficult to ``recognize'' a period, i.e. to notice that a complicated integral can be written in terms of known periods. 

The great mathematical relevance of this class of numbers comes from the fact that they can be dubbed, in a precise sense, as the \emph{numbers which come from ``geometry''}. Indeed, they can always be obtained from the comparison\footnote{After tensoring with $\mathbb{C}$, more generally considering ``relative cohomologies''.} between algebraic de Rham cohomology (related to algebraic differential forms) of an algebraic variety $X$ and Betti cohomology (related to topological cycles) of the complex points $X(\mathbb{C})$. One of the most important consequences of this fact is that period functions, and therefore Feynman integrals, satisfy special differential equations, called \emph{Picard-Fuchs equations}. Moreover, periods are strictly related to one of Grothendieck's deepest contributions to mathematics: the concept of \emph{motives}. Motives can be thought of as abstract linear algebra structures which encode the same kind of information as (all possible) cohomology theories of an algebraic variety, and which can conjecturally always be obtained as pieces of cohomologies of an actual (non-unique) algebraic variety. Therefore motives come equipped with a ``de Rham'' and a ``Betti'' vector space, isomorphic as $\mathbb{C}$-vector spaces, and from this one can define abstract analogues of periods, called \emph{motivic periods}, whose study has surprisingly deep consequences in the computation of Feynman integrals (motivic coaction, motivic f-alphabets..).

Originally, notably after the work of Broadhurst and Kreimer in the 1990's~\cite{BroadKreimer}, computations of Feynman integrals in the simplest (massless) cases seemed to assign a very special role to certain periods, first considered by Euler and then systematically studied by Zagier~\cite{ZagierMZV}, called \emph{multiple zeta values} (MZVs). They are defined as the absolutely convergent nested sums
\begin{equation}\label{MZV}
\zeta(k_1,\ldots ,k_r)=\sum_{0<v_1<\cdots <v_r}\frac{1}{v_1^{k_1}\cdots v_r^{k_r}},
\end{equation}
where $\textbf{k}:=(k_1,\ldots ,k_r)\in\mathbb{N}^r$ and $k_r\geq 2$. The rational vector space which they span will be denoted by $\mathcal{Z}$. One can easily demonstrate that this vector space is closed under the operation of taking products; in other words, $\mathcal{Z}$ is a $\mathbb{Q}$-algebra, which is conjecturally graded by the \emph{weight} $k_1+\cdots +k_r$. If we call $r$ the \emph{depth} of $\zeta(k_1,\ldots ,k_r)$, we can immediately see that depth-one MZVs are nothing but special values of the Riemann zeta function
\begin{equation*}
\zeta(s)=\sum_{v\geq 1}\frac{1}{v^s},
\end{equation*}
which are easily shown to be periods, because for each $k\geq 2$, using geometric series,
\begin{equation*}
\zeta(k)=\int_{[0,1]^k}\frac{dx_1\cdots dx_k}{1-x_1\cdots x_k}.
\end{equation*}
In fact, similar integral representations allow us to write all MZVs as periods. Physicists very quickly realized the convenience of trying to write down Feynman periods as MZVs, because of a nice explicit (conjectural) description of all relations in $\mathcal{Z}$, and especially because of the astonishing precision to which MZVs can be numerically approximated (thousands of digits within few seconds). The ubiquity of these numbers led to ask deep questions about their nature. Mathematicians realized that they are (geometric) periods of compactified moduli spaces of genus-zero Riemann surfaces with marked points $\overline{\mathfrak{M}}_{0,n}$ and that they can be seen as (real analogues of) motivic periods of a certain class of motives (mixed Tate motives over $\mathbb{Z}$, or $MT(\mathbb{Z})$), which are in some sense the simplest class of motives beyond those described by Grothendieck (called pure motives)~\cite{GonchManin, DeligneGonch}. A very good reason for the ubiquity of MZVs was given by Brown, who proved first that all periods of $\overline{\mathfrak{M}}_{0,n}$ belong to $\mathcal{Z}[2\pi i]$ and then that all periods of $MT(\mathbb{Z})$ belong to $\mathcal{Z}[1/2\pi i]$~\cite{BrownModuliSpace, BrownMTM}. Both viewpoints inspired the development of powerful techniques,
which can now be used to compute in a surprisingly short time vast classes of Feynman integrals. On the other side, these numbers are not sufficient to describe all interactions of particles, not even in the simplest models (scalar massless $\varphi^4$)~\cite{BelBros, BrownSchnetz}. 

Studying Feynman integrals beyond MZVs, studying periods of moduli spaces of Riemann surfaces beyond genus zero and studying motives beyond the mixed Tate case turned out to be different facets of the same problem. The study of superstring amplitudes seems to combine all these themes in a beautiful way.

\section{Superstring amplitudes}
\label{sec:3}

\vspace{1mm}\noindent

Scattering amplitudes in perturbative superstring theory can be approximated by a Feynman-like infinite sum of integrals over compactified moduli spaces of Riemann surfaces\footnote{More precisely, it should be super-Riemann surfaces, but this does not matter here.} with marked points. Each marked point represents a string state, and strings can be roughly divided between \emph{open strings} and \emph{closed strings}: this distinction translates into that between Riemann surfaces with and without boundaries, respectively, as shown in the figure below.

\begin{figure}[h!]
\begin{center}
\def\svgwidth{\columnwidth}
 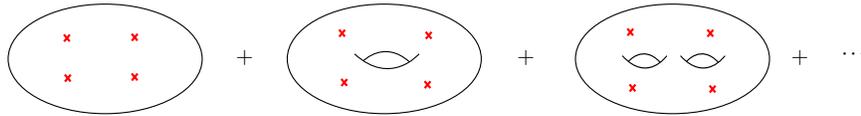
 \caption{Four closed strings: Riemann surfaces without boundaries}
    \end{center}
\end{figure}

\begin{figure}[h!]
\begin{center}
\def\svgwidth{\columnwidth}
 \scalebox{0.8}{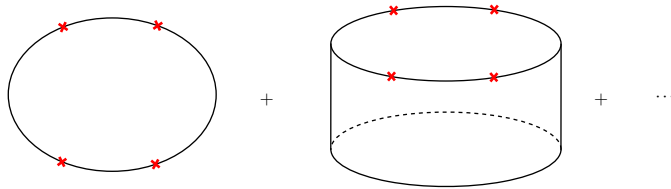}
 \caption{Four open strings: Riemann surfaces with boundaries, punctures lie on boundaries}
    \end{center}
\end{figure}

To each pair $(g,n)$ given by fixing the genus $g$ and the number of punctures $n$, one would like to associate a quantity $\textbf{A}^{\bullet}_{g,n}(\textbf{s})$: the genus--$g$ amplitude of~$n$ open or closed strings ($\bullet$ stands for $op$ or $cl$, respectively), i.e. a function of the \emph{Mandelstam variables} $\textbf{s}=(s_1,s_2,\ldots )$, which are complex numbers encoding the \emph{fundamental string tension} $\alpha^\prime$ and the momenta of the strings. It is important to mention that it is not clear how to define $\textbf{A}^{\bullet}_{g,n}(\textbf{s})$ for $g\geq 3$~\cite{DonagiWitten}. This paper is mainly concerned with the study of the mathematics associated with $(g,n)=(1,4)$. In order to speak of genus one it is useful to recall what is known about the mathematical structure of genus-zero amplitudes, which will be done in the next subsection. We would also like to mention that very encouraging progress has recently been made in genus two~\cite{DG2014, DGP, DGP2}.

\subsection{Genus zero}

Multiple zeta values (see eqn.~(\ref{MZV})) can be thought of as special values at $z=1$ of (one-variable) \emph{multiple polylogarithms}, defined for $|z|<1$ and $\textbf{k}\in\mathbb{N}^r$ by
\begin{equation*}
\text{Li}_{\textbf{k}}(z)=\sum_{0<v_1<\cdots <v_r}\frac{z^{k_r}}{v_1^{k_1}\cdots v_r^{k_r}}.
\end{equation*}
These functions can be analytically extended to holomorphic functions on the punctured Riemann sphere $\mathbb{P}^1_\mathbb{C}\setminus\{0,1,\infty\}$ by writing them as iterated integrals on paths from $0$ to $z$, but the analytic continuation depends on the homotopy class of the path. We say that they define \emph{multi-valued functions} on $\mathbb{P}^1_\mathbb{C}\setminus\{0,1,\infty\}$. Using their iterated-integral representation there is a natural way to make sense of their special values at $z=1$, which gives back our definition of MZVs whenever $k_r\geq 2$. Note that for instance $\text{Li}_1(z)=-\log(1-z)$ for $|z|<1$, which obviously extends to a multi-valued holomorphic function on the punctured Riemann sphere. There is a standard way to kill its monodromy and end up with a honest single-valued function: add its complex conjugate $-\log(1-\overline{z})$ and therefore get $-2Re(\log(1-z))=-\log|1-z|^2$. The price to pay is that we are giving up holomorphicity: we are left with a real analytic function on $\mathbb{P}^1_\mathbb{C}\setminus\{0,1,\infty\}$ (which extends continuously to $\mathbb{C}$). There is a natural generalization of this construction to all multiple polylogarithms $\text{Li}_{\,\textbf{k}}(z)$~\cite{BrownSVMPL}. We denote the single-valued analogues by $\mathcal{L}_{\textbf{k}}(z)$, we call them \emph{single-valued multiple polylogarithms} and we call \emph{single-valued multiple zeta values} their (regularized) special values at $z=1$:
\begin{equation*}
\zeta_{\rm sv}(k_1,\ldots ,k_r):=\mathcal{L}_{\,k_1,\ldots ,k_r}(1).
\end{equation*}
It turns out that these special values are contained in $\mathcal{Z}$~\cite{BrownSVMZV}, and actually form a much smaller sub-algebra, which we denote by $\mathcal{Z}^{\rm sv}$. For instance, one can prove that $\zeta_{\rm sv}(2k)=0$ and $\zeta_{\rm sv}(2k+1)=2\zeta(2k+1)$ for all $k\geq 1$.

As we have mentioned, string scattering amplitudes should be given by certain integrals over moduli spaces of Riemann surfaces. In practice, it is convenient to think of these integrals as divided into a first integration over all the possible positions of the marked points (string insertions) on a fixed (genus--$g$) Riemann surface, and then a second integral over all possible (complex structures of) genus--$g$ Riemann surfaces. In genus zero, there is only one possible Riemann surface: the Riemann sphere $\mathbb{P}^1_\mathbb{C}$ in the closed-string case and the unit disc in the open-string case, which requires boundaries. The integral needs to be invariant under the $\text{SL}_2(\mathbb{C})$-action ($\text{SL}_2(\mathbb{R})$ for open strings), so we can fix three points and finally think of our genus-zero $n$-point amplitudes as integrals over possible configurations of $n-3$ points on $\mathbb{P}^1_\mathbb{C}\setminus\{0,1,\infty\}$ in the closed-string case, or as (all possible) integrals of $n-3$ ordered points in the interval $[0,1]$ in the open-string case.

For what concerns the integrand, we will be even more sketchy. The general idea is that it depends on the Mandelstam variables and on two-variable functions, called \emph{propagators}, that are applied to pairs of marked points on the surface. Propagators are defined in terms of suitable \emph{Green functions}. The closed-string genus-zero propagator is $G^{cl}_0(z_1,z_2)=\log|z_1-z_2|^2$, while the open-string analogue is given by $G^{op}_0(x_1,x_2)=\log(x_1-x_2)$ for $0\leq x_2<x_1\leq 1$.

For instance, simplifying a bit, the genus-zero four-point open-string scattering amplitude (or Veneziano amplitude) is given by
\begin{equation}\label{g0op}
\int_0^1 \exp((s_1-1)\log x+(s_2-1)\log(1-x))\,dx
\end{equation}
and the closed-string analogue (Virasoro amplitude) is given by
\begin{equation}\label{g0cl}
-\frac{1}{2\pi i}\int_{\mathbb{P}^1_\mathbb{C}} \exp((s_1-1)\log|z|^2+(s_2-1)\log|z-1|^2)\,dzd\overline{z}
\end{equation}
People are interested in the asymptotic expansion of scattering amplitudes as the Mandelstam variables approach the origin. For instance, one finds that eqns.~(\ref{g0op}) and~(\ref{g0cl}) can be written, respectively, as
\begin{equation}\label{g0op2}
\frac{s_1+s_2}{s_1s_2}\,\exp\bigg(\sum_{n\geq 2}\frac{(-1)^n\zeta(n)}{n}\big(s_1^n+s_2^n-(s_1+s_2)^n\big)\bigg),
\end{equation}
\begin{equation}\label{g0cl2}
\frac{s_1+s_2}{s_1s_2}\exp\bigg(-2\sum_{n\geq 1}\frac{\zeta(2n+1)}{(2n+1)}(s_1^{2n+1}+s_2^{2n+1}-(s_1+s_2)^{2n+1})\bigg).
\end{equation}
These are actually the only cases where we are able to write down a closed formula for the asymptotic expansion in the Mandelstam variables. A simple crucial observation is the following: if we define a \emph{single-valued map}
\begin{equation*}
sv:\zeta(\textbf{k})\rightarrow \zeta_{\rm sv}(\textbf{k}),
\end{equation*}
and we extend it to the amplitude by leaving Mandelstam variables and rational numbers untouched, we conclude that the single-valued image of the open-string amplitude~(\ref{g0op2}) is precisely the closed-string amplitude~(\ref{g0cl2}). It is known (as a corollary of the work of Brown on the periods of $\mathfrak{M}_{0,n}$~\cite{BrownModuliSpace}) that, for any number $n$ of string insertions, the coefficients appearing in the asymptotic expansion of genus-zero open-string amplitudes are MZVs \cite{BSST}. Moreover, extending the trivial four-point observation made above by some ``motivic reasoning'' and many computer experiments, Schlotterer and Stieberger conjectured that the coefficients of the asymptotic expansion of genus-zero closed-string amplitudes should be given for any number of punctures by single-valued MZVs, and the closed-string amplitude should be (in a precise sense) the single-valued image of the open-string amplitude~\cite{ScSt, Stieb2014}.

\subsection{Genus one}

In genus one the structure of the moduli spaces gets more complicated, in particular in the open-string case, where we must take into account both oriented and non-oriented Riemann surfaces (cylinders and M\"obius strips). In the closed-string case, after fixing one marked point in order to insure translation invariance of the integrals, we are left with a first integration over~$n-1$ marked points on a fixed complex torus\footnote{It is well known that any punctured genus-one Riemann surface can be realized as a complex torus.} $\mathcal{E}_\tau=\mathbb{C}/(\tau\mathbb{Z}+\mathbb{Z})$, where $\tau\in\mathbb{H}$, and then a second integration over the (compactification of the) moduli space $\mathfrak{M}_{1,1}=\text{SL}_2(\mathbb{Z})\setminus\mathbb{H}$ of complex tori. This second integration is not trivial anymore, but in this paper we will only focus on the first integration, whose result will therefore depend on a parameter $\tau\in\mathbb{H}$. For what concerns the open-string case, not only we will focus on the first integral over the position of the insertions, but we furthermore restrict our analysis to the cylinder case with all insertions on one of the two boundary components. The complex structure of the cylinder will depend on a parameter $\tau\in i\mathbb{R}$ (one should think of a cylinder as ``half a torus'' whose complex parameter $\tau$ is purely imaginary). We refer to~\cite{BMRS} for more details about the general case. 

Once again, the integrands of these amplitudes are written in terms of Mandelstam variables and propagators, i.e. Green functions depending on two marked points on the surface. In the closed-string case the propagator is given by~\cite{GV}
\begin{equation}\label{closedprop}
G^{cl}_1(z_1,z_2;\tau)=-\log\bigg|\frac{\theta(z_1-z_2,\tau)}{\eta(\tau)}\bigg|^2+\frac{2\pi(Im(z_1)-Im(z_2))^2}{Im(\tau)},
\end{equation}
where the odd Jacobi $\theta$-function and the Dedekind $\eta$-function are defined for $z\in\mathbb{C}$, $\tau\in\mathbb{H}$, $q=\exp(2\pi i\tau)$ and $u=\exp(2\pi i z)$ by
\begin{equation}\label{Theta}
\theta(z,\tau)=q^{1/8}(u^{1/2}-u^{-1/2})\prod_{j\geq 1}(1-q^j)(1-q^ju)(1-q^ju^{-1}),
\end{equation}
\begin{equation*}
\eta(\tau)=q^{1/24}\prod_{j\geq 1}(1-q^j).
\end{equation*}
In the open-string case the propagator is usually defined as a ``regularized integral'': one can prove that for small $\epsilon$ and $z\in(0,1)$ there exists $K\in\mathbb{Z}_{\geq 0}$ such that
\begin{equation}\label{regintopen}
\int_\epsilon^{z} \frac{\theta^\prime (w,\tau)}{\theta(w,\tau)} dw = \sum_{k=0}^K g_k(z,\tau)\log^k(2\pi i\epsilon) + O(\epsilon).
\end{equation}
Then one defines\footnote{Note that we have changed the sign of the propagator w.r.t.~\cite{BMMS}, following~\cite{BSZ}.} the open-string genus-one propagator as~\cite{BMMS}
\begin{equation}\label{OpenPropImpl}
G^{op}_1(z_1,z_2;\tau)=-g_0(z_1-z_2,\tau).
\end{equation}
The reason in~\cite{BMMS} for giving such a non-explicit formula for the open-string propagator is that eqn.~(\ref{OpenPropImpl}) is the most convenient form to show that the open-string amplitude can be expressed in terms of \emph{elliptic multiple zeta values} (as defined by Enriquez~\cite{Enriquez}). It is however useful for our purposes to work out the regularized integral (\ref{regintopen}), which gives the explicit formula
\begin{equation}\label{OpenPropExpl}
G^{op}_1(z_1,z_2;\tau)=-\log\bigg(\frac{\theta(z_1-z_2,\tau)}{\eta^3(\tau)}\bigg).
\end{equation}
We will now define the genus-one Feynman-like integrals coming from string amplitudes which inspired the theory of modular and holomorphic graph functions.

The integral over the positions of the marked points on a fixed complex torus contributing to the four-point genus-one closed-string amplitude is given by \cite{GV}
\begin{multline}\label{4closedgenus1}
\int_{(\mathcal{E}_\tau)^3}\exp\big(s_1(G^{cl}_1(z_1,0;\tau)+G^{cl}_1(z_2,z_3;\tau))
+s_2(G^{cl}_1(z_2,0;\tau)+G^{cl}_1(z_1,z_3;\tau))\\
+s_3(G^{cl}_1(z_3,0;\tau)+G^{cl}_1(z_1,z_2;\tau))\big)\,d^2z_1d^2z_2d^2z_3,
\end{multline}
where we define $d^2z:=dzd\overline{z}/Im(\tau)$ and the Mandelstam variables must satisfy the relation $s_1+s_2+s_3=0$. 

On the other side, the integrals over the positions of the ordered marked points on one boundary of the cylinder topology, which contribute to the four-point genus-one open-string amplitude, are given by integrals like \cite{BMMS}
\begin{multline}\label{4opengenus1}
\int_{0\leq z_1\leq z_2\leq z_3\leq 1}\exp\big(s_1(G^{op}_1(z_1,0;\tau)+G^{op}_1(z_2,z_3;\tau))
+s_2(G^{op}_1(z_2,0;\tau)\\
+G^{op}_1(z_1,z_3;\tau))
+s_3(G^{op}_1(z_3,0;\tau)+G^{op}_1(z_1,z_2;\tau))\big)\,dz_1dz_2dz_3,
\end{multline}
where again we impose $s_1+s_2+s_3=0$. Different orderings of the $z_i$'s give rise to different contributions.

The theory of modular and holomorphic graph functions was born out of the study of the expansion of the integrals (\ref{4closedgenus1}) and (\ref{4opengenus1}), respectively, when the Mandelstam variables $s_i\rightarrow 0$.

\section{Modular graph functions}
\label{sec:4}

\vspace{1mm}\noindent

Let us consider a connected undirected graph $\Gamma$ with no self-edges, possibly with multiple edges connecting the same pair of vertices. If we choose a labelling $z_1,\ldots ,z_n$ of the $n$ vertices, then for $i<j$ we have $l_{i,j}$ edges between $z_i$ and $z_j$, with the total number of edges given by the \emph{weight} of the graph
\[
l:=\sum_{1\leq i<j\leq n}l_{i,j}.
\]
\begin{defn}[D'Hoker-Green-G\"urdo\u{g}an-Vanhove 2015]
Let $\Gamma$ be a graph as above. We define its \emph{modular graph function} as
\begin{equation}\label{modgraphfcts}
D_\Gamma(\tau)=\int_{(\mathcal{E}_\tau)^{n-1}}\prod_{1\leq i<j\leq n}G^{cl}_1(z_i,z_j;\tau)^{l_{i,j}}\,d^2z_1\cdots d^2z_{n-1},
\end{equation}
where $d^2z_i=dz_id\overline{z}_i/Im(z_i)$ and we have fixed $z_n\equiv 0$.
\end{defn}
It is obvious that this definition does not depend on the labelling. It will often be convenient to write a modular graph function $D_\Gamma(\tau)$ as $\textbf{D}[\Gamma]$, dropping the dependence of the function on $\tau$ and explicitly drawing the graph $\Gamma$, as in the following examples: $\Dcyc{G2}$, $\Dcyc{G211}$, $\Dcyc{G321}$, $\Dcyc{GC211}$ etc.. 

The definition of this class of functions originated from the study of the contribution to the genus-one four-point closed-string integral given by eqn.~(\ref{4closedgenus1})~\cite{GV, GRV, DGGV}. Indeed, if we expand the exponential in the integrand of~(\ref{4closedgenus1}) as a power series in the Mandelstam variables, the coefficients of this expansion are precisely the modular graph functions associated to graphs with at most four vertices. Modular graph functions associated to a higher number of vertices do not suffice to describe higher-point analogues of~(\ref{4closedgenus1}): one must introduce other functions called modular graph forms~\cite{DG}, but for the purpose of this paper it is enough to focus on modular graph functions. As anticipated in the introduction, recent considerations by Brown, contained in \cite{BrownNewClassI, BrownNewClassII}, seem to assign to this class of functions an important role in the study of mixed motives. We will come back to this in Section \ref{sec:6}.

So far we have only partially justified the origin of the name of these functions: it is clear that they are related to graphs, but something must be said about the word \emph{modular}. To begin with, one can show~\cite{ZagierBWR} that, if we denote $\Lambda_\tau=\mathbb{Z}+\tau\mathbb{Z}$, $\Lambda_\tau^*=\Lambda_\tau\setminus\{(0,0)\}$ and we define for $z\in\mathbb{C}$ the character on $\Lambda_\tau$ given for $\omega\in\Lambda_\tau^*$ by
\begin{equation*}
\chi_z(\omega):=\exp\Big(\frac{2\pi i(\bar{\omega}z-\omega\bar{z})}{\tau-\bar{\tau}}\Big),
\end{equation*}
then
\begin{equation}\label{svEllPol}
G(z_1,z_2;\tau)=\frac{Im(\tau)}{\pi}\sum_{\omega\in\Lambda_\tau^*}\frac{\chi_{z_1-z_2}(\omega)}{|\omega|^2}.
\end{equation}
Let us now introduce some more notation attached to the graph $\Gamma$. Choosing a labelling $z_1,\ldots ,z_n$ of the $n$ vertices induces an orientation on the edges: we orient them from $z_i$ to $z_j$ whenever $i<j$. Therefore we can construct the \emph{incidence matrix} 
\[
(\Gamma_{i,\alpha})_{\substack{1\leq i\leq n\\ 1\leq\alpha\leq l}}
\]
of $\Gamma$ by choosing any labelling $e_\alpha$ on the set of edges, and by setting $\Gamma_{i,\alpha}=0$ if $e_\alpha$ does not touch $z_i$, $\Gamma_{i,\alpha}=1$ if $e_\alpha$ is oriented away from $z_i$ and $\Gamma_{i,\alpha}=-1$ if $e_\alpha$ is oriented towards $z_i$. It is now not difficult to show that
\begin{equation}\label{sumrepmodfcts}
D_\Gamma(\tau)=\Big(\frac{Im(\tau)}{\pi}\Big)^l\sum_{\omega_1,\ldots ,\omega_l\in\Lambda_\tau^*}\prod_{\alpha=1}^l|\omega_\alpha|^{-2}\prod_{i=1}^n\delta\Big(\sum_{\beta=1}^l\Gamma_{i,\beta}\omega_\beta\Big),
\end{equation}
where $\delta(x)=1$ if $x=0$ and $\delta(x)=0$ otherwise. From eqn.~(\ref{sumrepmodfcts}) it is easy to see that modular graph functions are indeed modular invariant, i.e. $D_\Gamma(\gamma\tau)=D_\Gamma(\tau)$ for all $\gamma\in \mbox{SL}_2(\mathbb{Z})$, where $\gamma\tau$ is the M\"obius action of the modular group $\mbox{SL}_2(\mathbb{Z})$ on $\mathbb{H}$.

Let us now observe some simple consequences of the graphical nature of these functions. We call \emph{reducible} a graph $\Gamma$ such that the removal of a vertex would disconnect the graph, as in the figure below.

\begin{center}
\tikzpicture[scale=1]
\scope[xshift=-5cm,yshift=-0.4cm]
\draw[thick]   (0,0) node{$\bullet$} ..controls (1,0.7) .. (2,0) node{$\bullet$} ;
\draw[thick]   (0,0) node{$\bullet$} ..controls (1,-0.1) .. (2,0) node{$\bullet$} ;
\draw[thick]   (0,0) node{$\bullet$} ..controls (1,-0.7) .. (2,0) node{$\bullet$} ;
\draw[thick]   (2,0) node{$\bullet$} ..controls (3,0.7) .. (4,0) node{$\bullet$} ;
\draw[thick]   (2,0) node{$\bullet$} ..controls (3,-0.1) .. (4,0) node{$\bullet$} ;
\draw[thick]   (2,0) node{$\bullet$} ..controls (3,-0.7) .. (4,0) node{$\bullet$} ;
\endscope
\endtikzpicture
\end{center}
When a graph is reducible, one can prove from the sum representation~(\ref{sumrepmodfcts}) that the associated modular graph function factors into the product of the irreducible components. For instance, in the case of the figure above the modular graph function associated is~$\big(\Dcyc{G3}\big)^2$. Note also that the normalization of the Green function given by eqn.~(\ref{closedprop}) is chosen in such a way that
\begin{equation*}
\int_{\mathcal{E}_\tau}G_1^{cl}(z,0;\tau)dz=0.
\end{equation*} 
This, together with the factorization for reducible graphs, implies that $D_\Gamma(\tau)=0$ whenever there exists any edge whose removal would disconnect the graph (physicists would call such graphs \emph{one-particle reducible}). Below we have pictures of these situations:

\begin{center}
\tikzpicture[scale=1]
\scope[xshift=-5cm,yshift=-0.4cm]
\draw[thick]   (0,0) node{$\bullet$} ..controls (1,0.7) .. (2,0) node{$\bullet$} ;
\draw[thick]   (0,0) node{$\bullet$} ..controls (1,-0.1) .. (2,0) node{$\bullet$} ;
\draw[thick]   (0,0) node{$\bullet$} ..controls (1,-0.7) .. (2,0) node{$\bullet$} ;
\draw[thick]   (2,0) node{$\bullet$} ..controls (2.5,-0.1) .. (3,0) node{$\bullet$} ;
\draw[thick]   (3,0) node{$\bullet$} ..controls (4,0.7) .. (5,0) node{$\bullet$} ;
\draw[thick]   (3,0) node{$\bullet$} ..controls (4,-0.1) .. (5,0) node{$\bullet$} ;
\draw[thick]   (3,0) node{$\bullet$} ..controls (4,-0.7) .. (5,0) node{$\bullet$} ;
\draw[thick]   (7,0) node{$\bullet$} ..controls (8,0.7) .. (9,0) node{$\bullet$} ;
\draw[thick]   (7,0) node{$\bullet$} ..controls (8,-0.1) .. (9,0) node{$\bullet$} ;
\draw[thick]   (7,0) node{$\bullet$} ..controls (8,-0.7) .. (9,0) node{$\bullet$} ;
\draw[thick]   (9,0) node{$\bullet$} ..controls (9.5,-0.1) .. (10,0) node{$\bullet$} ;
\endscope
\endtikzpicture
\end{center}

Let us now give some concrete examples. For all graphs with~$n$ vertices along one cycle, as in the figure

\begin{center}
\tikzpicture[scale=0.6]
\scope[xshift=-5cm,yshift=-0.4cm]
\node at (-0.5,0.2) {$z_1$};
\draw[thick]   (0,0) node{$\bullet$} ..controls (0.5,0.865) .. (1,1.73) node{$\bullet$} ;
\node at (0.5,1.73) {$z_2$};
\draw[thick]   (1,1.73) node{$\bullet$}  ..controls (2,1.73) .. (3,1.73) node{$\bullet$} ;
\node at (3.5,1.73) {$z_3$};
\draw[thick] (3,1.73) node{$\bullet$}  ..controls (3.5,0.865) ..  (4,0) node{$\bullet$} ;
\node at (4.5,0.2) {$z_4$};
\draw[thick, dashed]   (4,0) node{$\bullet$} ..controls (3,-1.73) .. (1,-1.73) node{$\bullet$} ;
\node at (0.5,-1.73) {$z_n$};
\draw[thick]   (1,-1.73) node{$\bullet$} ..controls (0.5,-0.865) .. (0,0) node{$\bullet$} ;
\endscope
\endtikzpicture
\end{center}
we get by eqn.~(\ref{sumrepmodfcts}) that
\[
D_\Gamma(\tau)=\bigg(\frac{Im(\tau)}{\pi}\bigg)^n \sum_{\omega\in\Lambda_\tau^*}\frac{1}{|\omega|^{2n}}.
\]
This is precisely the definition of the special values at integers $n$ of the \emph{non-holomorphic Eisenstein series} $E(n,\tau)$. There are two well known results about non-holomorphic Eisenstein series which we want to underline, as some of their features are conjectured to extend to all modular graph functions:
\begin{itemize}
\item[1)] Let $\Delta_\tau:=4(Im(\tau))^2\frac{\partial^2}{\partial_\tau\partial_{\overline{\tau}}}$ be the hyperbolic Laplacian. Then
\begin{equation}\label{laplaceEis}
(\Delta_\tau-n(n-1))E(n,\tau)=0.
\end{equation}
\item[2)] Let $y=\pi\,Im(\tau)$, $B_k$ be the $k$-th Bernoulli number and $\sigma_j(k):=\sum_{d|k}d^j$. Then 
\begin{multline}\label{asympEis}
E(n,\tau)=\Big[(-1)^{n-1}\frac{B_{2n}}{(2n)!}(4y)^n+\frac{4(2n-3)!}{(n-2)!(n-1)!}\zeta(2n-1)(4y)^{1-n}\\
+\frac{2}{(n-1)!}\sum_{k\geq 1}k^{n-1}\sigma_{1-2n}(k)(q^k+\overline{q}^k)\sum_{m=0}^{n-1}\frac{(n+m-1)!}{m!(n-m-1)!}(4ky)^{-m}\Big].
\end{multline}
\end{itemize}
Laplace equations like~(\ref{laplaceEis}) were shown to hold for many other examples of modular graph functions. For instance, we have
\begin{equation}\label{diffeqG3}
(\Delta_\tau-6)\Dcyc{G3}=0
\end{equation}
as well as inhomogeneous equations like
\begin{equation*}
 (\Delta_\tau-6)
 \Dcyc{G2111}= \frac{86}{5} E(5,\tau) - 4 E(2,\tau)E(3,\tau) + \frac{ \zeta(5)}{10}
  \end{equation*}
and (infinitely) many others~\cite{DGV2015, Basu, DHokerKaidi}. It is generally believed that the algebra generated by modular graph functions over the ring $\mathcal{Z}$ of multiple zeta values should be closed under the action of~$\Delta_\tau$.\footnote{Recent indications suggest that considering only the action of the Laplace operator we lose some information, and it is instead better to consider the action of the \emph{Cauchy-Riemann derivative} $\nabla_\tau=2i(Im(\tau))^2\partial_\tau$ and of its complex conjugate $\overline{\nabla}_\tau$ \cite{DG, DHokerKaidi}.}

On the other side, the form of the asymptotic expansion (\ref{asympEis}) of non-holomorphic Eisenstein series generalizes to all modular graph functions. Indeed, as announced in the introduction, it was shown in~\cite{Zerb15} for up to four vertices and in~\cite{ZerbiniThesis} in the general case that the asymptotic expansion of a modular graph function has the following form:
\begin{thm}[Z. 2017]\label{Thm3}
Let $\mathcal{Z}_\infty$ be the $\mathbb{Q}$-algebra generated by all \emph{cyclotomic multiple zeta values}, i.e. all convergent series given for $k_1,\ldots ,k_r,N_1,\ldots ,N_r\in\mathbb{N}$ by
\begin{equation*}
\sum_{0<v_1<\cdots <v_r}\frac{e^{2\pi iv_1/N_1}\cdots e^{2\pi iv_r/N_r}}{v_1^{k_1}\cdots v_r^{k_r}}.
\end{equation*}
Then for a graph $\Gamma$ with weight $l$, setting $y=\pi \,Im(\tau)$, we have
\begin{equation*}
D_\Gamma (\tau)=\sum_{k=1-l}^l \sum_{m,n\geq 0} d_k^{(m,n)}(\Gamma)\, y^k q^m\overline{q}^n,
\end{equation*}
where $d_k^{(m,n)}(\Gamma)\in\mathcal{Z}_\infty$.
\end{thm}
The idea of the proof is to use the alternative representation of the propagator~\cite{GRV}
\begin{equation}\label{FourierProp2}
G_1(z,\tau)=2\pi\,Im(\tau)\overline{B}_2(r)+Q(z,\tau),
\end{equation}
where $\overline{B}_2(x)$ is the only 1-periodic continuous function coinciding with the second Bernoulli polynomial $B_2(x)$ in the interval $[0,1]$, $z=s+r\tau$ ($s,r\in [0,1]$) and
\begin{equation}\label{PPart}
Q(z,\tau)=\sum_{\substack{m\in\mathbb{Z}\setminus\{0\}\\k\in\mathbb{Z}}}\frac{\exp(2\pi im((k+r)Re(\tau)+s))}{|m|}e^{-2\pi\,Im(\tau)|m||k-r|}.
\end{equation}
The integral can be thought of as an integral over $r\in [0,1]$ and $s\in [0,1]$. One can therefore substitute $\overline{B}_2(r)$ with the polynomial $B_2(r)=r^2-r+1/6$. Using the binomial theorem to compute the powers of the propagator in terms of $B_2(r)$ and $Q(z,\tau)$ and interchanging integration with all summations introduced by the functions~$Q$, one is left with integrals that evaluate to an asymptotic expansion of the form
\begin{equation*}
\sum_{k=-l}^l\, \sum_{\substack{u\in\mathbb{Z}\\v\geq 1}} \delta_{k}^{(u,v)}(\Gamma)\,y^k e^{2\pi i uRe(\tau)}e^{-2\pi vIm(\tau)},
\end{equation*}
where $\delta_{k}^{(u,v)}(\Gamma)$ are certain explicitly determined complex numbers given in terms of complicated multiple sums and $u, v$ are subject to certain explicit constraints. To conclude the proof, one must first of all do the relatively easy exercise of showing that $\delta_{-l}^{(u,v)}(\Gamma)=0$, that $v\geq |u|$ and that $u\equiv v$ mod~2: we need the last two conditions because $e^{2\pi i uRe(\tau)}e^{-2\pi vIm(\tau)}=q^m \overline{q}^n$ with $m=(u+v)/2$ and $n=(v-u)/2$, and we want to get $m,n\in\mathbb{Z}_{\geq 0}$. The last step consists in proving that the coefficients are cyclotomic MZVs, but this is non-elementary and relies on a result of Terasoma~\cite{TerasomaConical}. The reader can find the details in~\cite{ZerbiniThesis}. However, later we will see that this is (conjecturally) not the best possible result.

The main contribution to this asymptotic expansion as $y\rightarrow \infty$ is given by the Laurent polynomial
\begin{equation}\label{dLaurent}
\textbf{d}[\Gamma]:=\sum_{k=1-l}^l d_k^{(0,0)}(\Gamma)\,y^k,
\end{equation}
and great effort was spent on its computation, justified by the fact that all known differential and algebraic relations among modular graph functions can be predicted from its knowledge\footnote{It is however believed that this should not always be true.}. The first computations were made in \cite{GV} and \cite{GRV}. In particular, in the appendix of \cite{GRV} Zagier gave a general formula in terms of MZVs for $\textbf{d}[\Gamma]$ when $\Gamma$ is a ``banana'' graph consisting only on two vertices and $l$ edges between them. Later on, Zagier also proved that one can be more precise and write $\textbf{d}[\Gamma]$ for all banana graphs in terms of odd Riemann zeta values $\zeta(2k+1)$ \cite{ZagierStrings}, just like the case of non-holomorphic Eisenstein series. Computations by hands for graphs with three and four vertices seemed to confirm the appearence of odd Riemann zeta values only \cite{GRV}. A more systematic computation of the Laurent polynomials $\textbf{d}[\Gamma]$, however, revealed that already for graphs with three vertices one can get MZVs of higher depth \cite{Zerb15}. For instance\footnote{There is a typo in the coefficient of $y^{-4}$ in the corresponding formula in \cite{Zerb15}.},
\begin{multline*}
\smallDcyc{G511}=\frac{62}{10945935}y^7+\frac{2}{243}\zeta(3)y^4+\frac{119}{324}\zeta(5)y^2 +\frac{11}{27}\zeta(3)^2y+\frac{21}{16}\zeta(7)\\
+\frac{46}{3}\frac{\zeta(3)\zeta(5)}{y}+\frac{7115\zeta(9)-3600\zeta(3)^3}{288y^2}+\frac{1245\zeta(3)\zeta(7)-150\zeta(5)^2}{16y^3}\\
+\frac{288\zeta(3)\zeta(3,5)-288\zeta(3,5,3)-5040\zeta(5)\zeta(3)^2-9573\zeta(11)}{128y^4}\\
+\frac{2475\zeta(5)\zeta(7)+1125\zeta(9)\zeta(3)}{32y^5}-\frac{1575}{32}\frac{\zeta(13)}{y^6}.
\end{multline*}
The key observation made in \cite{Zerb15} is that one can rewrite
\begin{multline}\label{Dweight7}
\smallDcyc{G511}=\frac{62}{10945935}y^7+\frac{1}{243}\zeta_{\rm sv}(3)y^4+\frac{119}{648}\zeta_{\rm sv}(5)y^2 +\frac{11}{108}\zeta_{\rm sv}(3)^2y+\frac{21}{32}\zeta_{\rm sv}(7)\\
+\frac{23}{6}\frac{\zeta_{\rm sv}(3)\zeta_{\rm sv}(5)}{y}+\frac{7115\zeta_{\rm sv}(9)-900\zeta_{\rm sv}(3)^3}{576y^2}+\frac{1245\zeta_{\rm sv}(3)\zeta_{\rm sv}(7)-150\zeta_{\rm sv}(5)^2}{64y^3}\\
-\frac{288\zeta_{\rm sv}(3,5,3)+1620\zeta_{\rm sv}(5){\zeta_{\rm sv}(3)}^2+9573\zeta_{\rm sv}(11)}{256y^4}\\
+\frac{2475\zeta_{\rm sv}(5)\zeta_{\rm sv}(7)+1125\zeta_{\rm sv}(9)\zeta_{\rm sv}(3)}{128y^5}-\frac{1575}{64}\frac{\zeta_{\rm sv}(13)}{y^6}.
\end{multline}
This observation extends to all graphs computed so far. Moreover, it also seems to extend to all other Laurent polynomials appearing in the expansion, and it is consistent with the genus-zero case. This led to the following conjecture (initially stated only in the four-point case \cite{Zerb15}):
\begin{conj}[Z. 2015]\label{Conjsvmzv}
The coefficients $d_k^{(m,n)}(\Gamma)$ given by Theorem \ref{Thm3} belong to the algebra $\mathcal{Z}^{\rm sv}$ of single-valued MZVs.
\end{conj}
It is important to remark that this conjecture is much stronger than the statement of Theorem \ref{Thm3}, which does not even imply that the coefficients of the asymptotic expansion are MZVs. Arguments supporting or proving special cases of this conjecture were given in \cite{ZagierStrings, DGGV, DHokDuke}.

We conclude this section by mentioning that an explicit computation of (part of) the asymptotic expansions, together with the differential relations among modular graph functions, allow us to prove algebraic relations such as
\begin{equation*}
 \Dcyc{G3} =\Dcyc{G111} + \zeta(3)  \, ,
\end{equation*}
which can be deduced using the differential equations  (\ref{laplaceEis}) and (\ref{diffeqG3}) and the knowledge of $\smallDcyc{G3}$ \cite{DGV2015}.\footnote{This identity was first proven by Zagier by a complicated direct computation (private communication).}

\section{Holomorphic graph functions}
\label{sec:5}

\vspace{1mm}\noindent

Let us consider a graph $\Gamma$ as in the beginning of Section \ref{sec:4}. Recall that in genus zero we had (special values of) holomorphic multi-valued multiple polylogarithms on the open-string side and real-analytic single-valued multiple polylogarithms on the closed-string side. These two sides were related by the $sv$ map. Since modular graph functions are real-analytic modular functions, i.e. single-valued on $\mathfrak{M}_{1,1}=\mbox{SL}_2(\mathbb{Z})\setminus \mathbb{H}$, in order to make an analogy with genus zero we would like to associate to any graph $\Gamma$ a holomorphic multi-valued function on $\mathfrak{M}_{1,1}$, together with a map going from one space to the other, which we would like to call $esv$ (elliptic single-valued map). Moreover, we would like these holomorphic graph functions to arise from open-string amplitudes, as this would make the analogy with genus zero satisfactory also from the string-theory viewpoint.

A suggestion to achieve all this was given in \cite{BSZ}, where not just one but two kinds of holomorphic graph functions, related to each other by a modular transformation, were defined. In order to give the definition, we first need to introduce a modified version of the open-string propagator $G^{op}_1(z_i,z_j;\tau)$:
\begin{equation}\label{NewProp}
P(z_1,z_2;\tau):=G^{op}_1(z_i,z_j;\tau)-2\log(\eta(\tau))+\frac{i\pi\tau}{6}+\frac{i\pi}{2}
\end{equation}
\begin{defn}[Br\"odel-Schlotterer-Z. 2018]
Let $z_1,\ldots ,z_n\in [0,1]$ be the vertices of a graph $\Gamma$, let us fix $z_n\equiv 0$ and for $i<j$ let us denote the number of edges between $z_i$ and $z_j$ by $l_{i,j}$. For such~$\Gamma$ we define its \emph{A-cycle graph function} as
\begin{equation}\label{A-cyclegraphfct}
A_\Gamma(\tau)=\int_{[0,1]^{n-1}}\prod_{1\leq i<j\leq n}P(z_i,z_j;\tau)^{l_{i,j}}\,dz_1\cdots dz_{n-1},
\end{equation}
and its \emph{B-cycle graph functions} as
\begin{equation}\label{B-cyclegraphfct}
B_\Gamma(\tau)=A_\Gamma(-1/\tau).
\end{equation}
\end{defn}
We will often make use also of the alternative notations $\textbf{A}[\Gamma]$ and $\textbf{B}[\Gamma]$. Note that in general the integral~(\ref{A-cyclegraphfct}) may diverge. In this case we consider instead its $\varepsilon$-regularized version given by the same procedure already explained in the definition of $G_1^{op}(z_1,z_2;\tau)$ (see eqn.~(\ref{regintopen})).

Without going into details, A-cycle graph functions can be thought of as restrictions of the integral on a torus which defines modular graph functions, given by eqn.~(\ref{modgraphfcts}), to the ``A-cycle''~$[0,1]$ of the torus, while B-cycle graph functions as restrictions of (\ref{modgraphfcts}) to the ``B-cycle''~$[0,\tau]$. This is a first reason why it is worth introducing both notions, even though one can be obtained from the other by a simple modular transformation. We will see later that another very important reason is given by their radically different asymptotic expansions. However, since eqn.~(\ref{B-cyclegraphfct}) implies that these two classes of functions share many properties, we will sometimes refer to both of them as just \emph{holomorphic graph functions}: indeed, by the definition~(\ref{NewProp}) of $P(z_1,z_2;\tau)$, both A-cycle and B-cycle graph functions are holomorphic functions of $\tau\in\mathbb{H}$.

Let us now see how holomorphic graph functions are related to the four-point open-string integral~(\ref{4opengenus1}). First of all, note that adding any $(z_i,z_j)$-independent term to the open-string propagator $G^{op}_1(z_i,z_j;\tau)$ does not modify the open-string integral~(\ref{4opengenus1}), because of the kinematic condition $s_1+s_2+s_3=0$ on the Mandelstam variables. Thus considering the modified propagator $P(z_1,z_2;\tau)$ does not affect the amplitude. Let us now define the \emph{abelianization} of~(\ref{4opengenus1}) as
\begin{multline}\label{abelian}
\int_{[0,1]^3}\exp\big(s_1(P(z_1,0;\tau)+P(z_2,z_3;\tau))
+s_2(P(z_2,0;\tau)+P(z_1,z_3;\tau))\\
+s_3(P(z_3,0;\tau)+P(z_1,z_2;\tau))\big)\,dz_1dz_2dz_3.
\end{multline}
This is nothing but the sum over all possible orderings of the open-string positions on the cylinder's boundary~$[0,1]$ of the integrals of the kind~(\ref{4opengenus1}); it is called ``abelianization'' because it would correspond in physics to the amplitude of so-called abelian particle states, like photons, which are however not included in superstring theories. Even though the integral~(\ref{abelian}) is not ``physical'', it is clearly related to the open-string amplitude. Expanding~(\ref{abelian}) as a power series in the Mandelstam variables and allowing $\tau\in\mathbb{H}$, by a computation which is completely similar to that of the closed-string case we find that the coefficients are given by A-cycle graph functions (associated to graphs with at most four vertices), hence the connection between holomorphic graph functions and genus-one open-string amplitudes.

The reason for the normalization~(\ref{NewProp}) of the propagator is given by the fact that
\begin{equation}\label{vanishP}
\int_0^1 P(z,0;\tau)\, dz=0,
\end{equation}
which implies, as in the case of modular graph functions, that holomorphic graph functions vanish identically whenever they are associated to graphs that can be disconnected by removing one edge. The proof of eqn.~(\ref{vanishP}) using the $\varepsilon$-regularization procedure~(\ref{regintopen}) is left as an exercise to the reader.

In order to talk about non-trivial examples of holomorphic graph functions, we should first of all recall the observation, made in~\cite{BMMS}, that the coefficients of the power series expansion in the Mandelstam variables of the genus-one open-string integral~(\ref{4opengenus1}), and therefore also A-cycle graph functions\footnote{This is only true for graphs with at most four vertices, but there is an obvious $n$-point version of the integral~(\ref{abelian}) whose coefficients, i.e. all possible graph functions, must be combinations of A-elliptic MZVs.}, can be written in terms of \emph{A-elliptic multiple zeta values}. These are holomorphic functions on~$\mathbb{H}$ defined by Enriquez which generalize MZVs to genus one~\cite{Enriquez}. The~``A'' in their name comes from the fact that A-elliptic MZVs are given by certain iterated integrals over the A-cycle~$[0,1]$ of a torus $\mathcal{E}_\tau$. There exists also a B-cycle version given by iterated integrals over $[0,\tau]$: they are called \emph{B-elliptic multiple zeta values} and they are related to their A-cycle counterparts by the modular transformation $S: \tau\rightarrow -1/\tau$. Giving the definition of elliptic MZVs is not necessary here; we will just recall in the next paragraphs the properties needed in our presentation of holomorphic graph functions, and refer the interested reader to~\cite{Enriquez, MatthesThesis, ZerbiniThesis}.

The main property that we will need is that elliptic MZVs can be written as (iterated) integrals of Eisenstein series. We recall that Eisenstein series are given for even $k\geq 4$ by 
\begin{equation*}
G_{k}(\tau)=\sum_{(r,s)\neq (0,0)}\frac{1}{(r+s\tau)^{k}}=2\zeta(k)+\frac{2(2\pi i)^{k}}{(k-1)!}\sum_{m,n\geq 1}n^{k-1}q^{mn}.
\end{equation*}
They are weight $k$ modular forms w.r.t. the modular group $\mbox{SL}_2(\mathbb{Z})$, i.e. they are holomorphic functions on $\mathbb{H}\cup i\infty$ such that $G_{2k}(\tau)|_{2k}\,\gamma=G_{2k}(\tau)$, where the weight~$k$ right action $|_{k}\,\gamma$ of the modular group is defined for $\gamma=\left( \begin{array}{ccc}
a & b \\
c & d \end{array} \right)\in\mbox{SL}_2(\mathbb{Z})$ by
\begin{equation*}
f(\tau)|_{k}\,\gamma=(c\tau+d)^{-k}f\bigg(\frac{a\tau+b}{c\tau+d}\bigg).
\end{equation*}
Setting $G_0(\tau):=-1$, \emph{iterated Eisenstein integrals} are defined in~\cite{BMS} for $k_1,\ldots ,k_r\in\{0,4,6,8,\ldots \}$\footnote{Here we deviate from \cite{BMS} and we prefer to exclude the quasi-modular form $G_2(\tau)$.} by the recursive formula
\begin{equation}\label{itEisInt}
\mathcal{E}(k_1,\ldots ,k_r;\tau)=\int_{\tau}^{\overrightarrow{\textbf{1}}_{\infty}} \frac{G_k(z)}{(2\pi i)^{k-1}}\mathcal{E}(k_1,\ldots ,k_{r-1};z)\,dz,
\end{equation}
where following~\cite{MMV} we define for $f(\tau)=\sum_{j\geq 0}a_j(\tau) q^j$ and $a_j(\tau)\in\mathbb{C}[\tau]$\footnote{Since $\mathbb{H}$ is simply connected, we can choose arbitrary paths from~$\tau^\prime$ to~$i\infty$ and from~$0$ to~$\tau^\prime$.}
\begin{equation}\label{regul}
\int_{\tau^\prime}^{\overrightarrow{\textbf{1}}_{\infty}}f(\tau)\,d\tau:=\int_{\tau^\prime}^{i\infty}\sum_{j\geq 1}a_j(\tau) q^j\,d\tau-\int_0^{\tau^\prime} a_0(\tau)\,d\tau.
\end{equation}

Writing down A-cycle graph functions in terms of the iterated integrals $\mathcal{E}(\textbf{k};\tau)$ is very convenient, because all algebraic relations among the $\mathcal{E}(\textbf{k};\tau)$'s are known~\cite{MatthesLoschakSchneps} and because their asymptotic expansion can be explicitly worked out~\cite{BMS, BSZ}. Moreover, most importantly for us, this is a good viewpoint if one is interested in the modular behaviour of A-cycle graph functions, for instance in order to compute B-cycle graph functions~\cite{BSZ}. 

Let us now consider a simple example: one can show by direct computation that
\begin{eqnarray}
\Acyc{G2}&=&\frac{(\pi i\tau)^2}{60}+\frac{\zeta(2)}{2}-6\mathcal{E}(4,0;\tau)\label{firstline}\\
&=&\frac{\zeta(2)}{2}+2q+\frac{9}{2}q^2+\ldots\label{secondline}
\end{eqnarray}
The fact that the first term of the right-hand side of~(\ref{firstline}) disappears in~(\ref{secondline}) is not an accident: it is coherent with the fact, proven by Enriquez~\cite{Enriquez}, that A-elliptic MZVs (and therefore A-cycle graph functions) admit an asymptotic expansion
\begin{equation}\label{AsymptA}
\sum_{j\geq 0}a_j q^j,
\end{equation}
where $a_j\in\mathcal{Z}[(2\pi i)^{-1}]$.\footnote{While for A-elliptic MZVs we know that inverting $2\pi i$ is necessary, we suspect that no inverse powers of $2\pi i$ should appear in the asymptotic expansion of A-cycle graph functions.} As we have mentioned, writing down A-cycle graph functions in terms of iterated Eisenstein integrals is the key to compute B-cycle graph functions. In order to get $\Bcyc{G2}$, all we need to do is to compute $\mathcal{E}(4,0;-1/\tau)$. Let us see some details of this computation, which nicely illustrates various features of the general case. Directly from the definition, one gets
\begin{equation*}
\mathcal{E}(4,0;\tau)=\frac{1}{(2\pi i)^2}\int_\tau^{\overrightarrow{\textbf{1}}_{\infty}}(\tau-z)G_4(z)\,dz.
\end{equation*}
Therefore, making use of the change of variables $z\rightarrow -1/z$ and of the modular properties of $G_4(\tau)$, we can write
\begin{equation*}
\mathcal{E}(4,0;-1/\tau)=\frac{1}{(2\pi i)^2}\int_{-1/\tau}^{\overrightarrow{\textbf{1}}_{\infty}}\Big(-\frac{1}{\tau}-z\Big)G_4(z)\,dz=\frac{\tau^{-1}}{(2\pi i)^2}\int_{\tau}^{S\overrightarrow{\textbf{1}}_{\infty}}z(\tau-z)G_4(z)\,dz,
\end{equation*}
where $S\overrightarrow{\textbf{1}}_{\infty}$ is the image of the (tangential base point at the) cusp under the modular transformation $S:\tau\rightarrow -1/\tau$: one should think of it as the point $0$, together with additional informations about the regularization of the integral, similar to those of eqn.~(\ref{regul})~\cite{MMV}. Since these integrals are all homotopy invariant, we can choose a path which passes through the cusp $\overrightarrow{\textbf{1}}_{\infty}$ and split the integral as
\begin{equation}\label{chipichipi}
\mathcal{E}(4,0;-1/\tau)=\frac{\tau^{-1}}{(2\pi i)^2}\bigg(\int_\tau^{\overrightarrow{\textbf{1}}_{\infty}}z(\tau-z)G_4(z)\,dz-\int_{S\overrightarrow{\textbf{1}}_{\infty}}^{\overrightarrow{\textbf{1}}_{\infty}}z(\tau-z)G_4(z)\,dz\bigg)
\end{equation}
Using the fact that
\begin{eqnarray*}
2\mathcal{E}(4,0,0;\tau)&=&\frac{1}{2\pi i}\int_\tau^{\overrightarrow{\textbf{1}}_{\infty}}(\tau-z)^2G_4(z)\,dz\\
&=&\frac{1}{2\pi i}\int_\tau^{\overrightarrow{\textbf{1}}_{\infty}}\tau(\tau-z)G_4(z)\,dz-\frac{1}{2\pi i}\int_\tau^{\overrightarrow{\textbf{1}}_{\infty}}z(\tau-z)G_4(z)\,dz,
\end{eqnarray*}
one immediately sees that the first integral of eqn.~(\ref{chipichipi}) can be written as $\mathcal{E}(4,0;\tau)-(\pi i\tau)^{-1}\mathcal{E}(4,0,0;\tau)$. The evaluation of the second integral of eqn.~(\ref{chipichipi}) can be done using the functional equation of the L-functions associated to Eisenstein series: the computation is completely similar to that of period polynomials of Eisenstein series (see~\cite{ZagierPeriodsJacobi}). This is not a coincidence, because $\mathcal{E}(4,0,0;\tau)$ is essentially an Eichler integral of $G_4(\tau)$. The general theory of \emph{iterated Eichler integrals of modular forms}, developed by Manin and Brown \cite{ManinItInt, MMV}, provides the tools to understand the modular behaviour of the functions $\mathcal{E}(\textbf{k},\tau)$. It is crucial to remark that not all the $\mathcal{E}(\textbf{k},\tau)$'s can be written in terms of iterated Eichler integrals, and indeed not all of them share the same nice modular behaviour of our example $\mathcal{E}(4,0;\tau)$: it is an instructive exercise to see how the steps of the computations of $\mathcal{E}(4,0;-1/\tau)$ cannot be repeated for $\mathcal{E}(4,0,0,0;-1/\tau)$. Fortunately, a result by Brown implies that A-cycle graph functions can always be written in terms of the ``good $\mathcal{E}(\textbf{k},\tau)$'s'', i.e. those which are iterated Eichler integrals~\cite{BrownNewClassII}. We refer to~\cite{BSZ} for a detailed analysis of the relationship between the $\mathcal{E}(\textbf{k},\tau)$'s and iterated Eichler integrals \`a la Manin-Brown. The final result is that
\begin{align}
\Bcyc{G2}&=\frac{\zeta(2)}{3}-\frac{\zeta(3)}{T}-\frac{3\zeta(4)}{2T^2}-6\mathcal{E}(4,0;\tau)+\frac{6}{T}\mathcal{E}(4,0,0;\tau)\notag\\
&= \frac{T^2}{180}+\frac{\zeta(2)}{3}-\frac{\zeta(3)}{T}-\frac{3\zeta(4)}{2T^2}+2\bigg(1-\frac{1}{T}\bigg)q+\ldots \,,\label{Bexample}
\end{align}
where $T:=\pi i\tau$. We can already see from this simple example that the asymptotic expansion of B-cycle graph functions is substantially different from that of their A-cycle counterpart. In particular, B-cycle graph functions are not invariant under the transformation $\tau\rightarrow \tau+1$. Far from being an issue, this is actually the main reason behind the introduction of B-cycle graph functions in~\cite{BSZ}: an expansion like~(\ref{Bexample}), unlike eqn.~(\ref{AsymptA}), reminds the asymptotic behaviour of modular graph functions. We will indeed see in the next section that this simple observation has astonishing consequences. The main result of this section is the following refinement of a formula stated in~\cite{BSZ} for the asymptotic expansion of B-cycle graph functions.
\begin{thm}\label{Thm4}
For a graph $\Gamma$ with weight~$l$, setting $T=\pi i\tau$, we have
\begin{equation*}
B_\Gamma (\tau)=\sum_{k=-l}^l\,\sum_{m\geq 0} b_k^{(m)}(\Gamma)\, T^k q^m,
\end{equation*}
where $b_k^{(m)}(\Gamma)\in\mathcal{Z}$.
\end{thm}
\textbf{Proof.} It was demonstrated in~\cite{BSZ} that
\begin{equation}
B_\Gamma (\tau)=\sum_{k=-K}^K\,\sum_{m\geq 0} b_k^{(m)}(\Gamma)\, T^k q^m
\end{equation}
for some $K\in\mathbb{N}$, where $b_k^{(m)}(\Gamma)\in\mathcal{Z}$, so all we need to prove is that we can choose~$K=l$. In order to do this, let us recall the modular properties of~$\theta$ and~$\eta$:
\begin{align*}
\theta(z/\tau,-1/\tau)&=-i\sqrt{-i\tau}\,\exp\bigg(\frac{2\pi iz^2}{2\tau}\bigg)\theta(z,\tau),\\
\eta(-1/\tau)&=\sqrt{-i\tau}\,\eta(\tau).
\end{align*}
Using these transformations together with eqns~(\ref{NewProp}) and~(\ref{Theta}) we deduce that
\begin{multline}
P(z_i,z_j;-1/\tau)=-\frac{\pi i}{6\tau}+\frac{\pi i}{2}\\
-\log\bigg(i e^{\pi i\tau(z_i-z_j)^2}q^{1/12}(\tilde{u}_{ij}^{-1/2}-\tilde{u}_{ij}^{1/2})\prod_{n\geq 1}(1-\tilde{u}_{ij}q^n)(1-\tilde{u}_{ij}^{-1}q^n)\bigg),
\end{multline}
where $\tilde{u}_{ij}:=\exp(2\pi i\tau(z_i-z_j))$. Therefore, setting $z_{ij}:=z_i-z_j$, we can write 
\begin{equation}\label{P=L+S}
P(z_i,z_j;-1/\tau)=L(z_{ij};\tau)+S(z_{ij};\tau),
\end{equation}
where we define for $T=\pi i\tau$ and $\tilde{u}=\exp(2\pi i\tau z)$
\begin{equation}\label{L}
L(z,\tau)=-T\Big(z^2-z+\frac{1}{6}\Big)+\frac{\zeta(2)}{T},
\end{equation}
\begin{equation}\label{S}
S(z,\tau)=\sum_{m\geq 1}\frac{\tilde{u}^m}{m}+\sum_{n,m\geq 1}\frac{\tilde{u}^mq^{nm}}{m}+\sum_{n,m\geq 1}\frac{\tilde{u}^{-m}q^{nm}}{m}.
\end{equation}
In order to compute B-cycle graph functions, we need to take powers of the propagator. We have
\begin{align}
&P(z_i,z_j;-1/\tau)^l=(L(z_{ij};\tau)+S(z_{ij};\tau))^l=\sum_{r+s=l}\frac{l!}{r!s!}L(z_{ij},\tau)^rS(z_{ij},\tau)^s\notag\\
&=\sum_{a+b+c+d+s=l}\frac{l!}{a!b!c!d!s!}\frac{(-1)^{a+c}}{6^c}\zeta(2)^dz_{ij}^{2a+b}T^{a+b+c-d}S(z_{ij},\tau)^s. \label{chasingpowers}
\end{align}
Therefore, if for instance we want to know the asymptotic expansion of 
\begin{equation*}
\int_0^1P(z,0;-1/\tau)^l\,dz,
\end{equation*} 
i.e. 2-point B-cycle graph functions, we are left with computing integrals of the kind
\begin{equation*}
\int_0^1 z^{2a+b}S(z,\tau)^s\,dz.
\end{equation*}
If $s=0$ this integral evaluates to a rational numbers; thus we get a Laurent polynomial with powers ranging from~$-l$ (when~$d=l$) to~$l$ (when~$d=0$). Now let $s\geq 1$. Since for all $\alpha\in\mathbb{Z}$
\begin{equation}\label{2ndContr}
\int_0^1 z^{2a+b}e^{2\pi i\alpha\tau z}\,dz=\frac{(2a+b)!}{(-2\pi i\alpha\tau)^{2a+b+1}}-\sum_{j=0}^{2a+b}\frac{(2a+b)_j}{(-2\pi i\alpha\tau)^{j+1}}e^{2\pi i\alpha\tau},
\end{equation}
where $(k)_j:=k(k-1)\cdots (k-j+1)$ is the descending Pochhammer symbol, exchanging integration and summations in~(\ref{S}) we get negative contributions to the powers of~$T$ ranging from~$1$ to~$2a+b+1$. The lowest possible power that arises from these contributions is then given, because of eqn.~(\ref{chasingpowers}), by
\begin{equation*}
a+b+c-d-2a-b-1=c-a-d-1\geq -l+s-1\geq -l.
\end{equation*}
This concludes the proof when the graphs~$\Gamma$ has only  two vertices. The proof for the general case goes along the same lines and is left to the reader.\\
$\square$

\section{The \emph{esv} conjecture}
\label{sec:6}

\vspace{1mm}\noindent

The main contribution to the asymptotic expansion of B-cycle graph functions $\textbf{B}[\Gamma]$ when $Im(\tau)\rightarrow \infty$ is given by the first Laurent polynomial
\begin{equation}\label{bLaurent}
\textbf{b}[\Gamma]:=\sum_{k=-l}^l b_k^{(0)}(\Gamma)\,T^k.
\end{equation}
For instance, in the case of the weight $l=2$ B-cycle graph function~$\Bcyc{G2}$ we deduce by eqn.~(\ref{Bexample}) that
\begin{equation*}
\smallBcyc{G2}=\frac{T^2}{180}+\frac{\zeta(2)}{3}-\frac{\zeta(3)}{T}-\frac{3\zeta(4)}{2T^2}.
\end{equation*}
On the other side, we have mentioned in Section~\ref{sec:4} that $\Dcyc{G2}=E(2,\tau)$, and therefore by eqn.~(\ref{asympEis}) we have
\begin{equation*}
\smallDcyc{G2}=\frac{y^2}{45}+\frac{\zeta(3)}{y}.
\end{equation*}
Let us now define a map $\mathcal{Z}[T]\rightarrow \mathcal{Z}^{\rm sv}[y]$ by sending $\zeta(\textbf{k})\rightarrow \zeta_{\rm sv}(\textbf{k})$ and $T\rightarrow -2y$. We call this map~\emph{esv}. Using the fact that $\zeta_{\rm sv}(2k)=0$ and $\zeta_{\rm sv}(2k+1)=2\zeta(2k+1)$, one can easily check that $esv(\smallBcyc{G2})=\smallDcyc{G2}$. This was the starting point of an extensive series of computations, which led to the following apparently surprising statement \cite{BSZ}:
\begin{conj}[Br\"odel-Schlotterer-Z. 2018] 
For all graphs~$\Gamma$ we have
\begin{equation}\label{esvconj}
esv(\textbf{b}[\Gamma])=\textbf{d}[\Gamma]
\end{equation}
\end{conj}
For the rest of the paper we will refer to this statement as the \emph{esv conjecture}. We would like to remark, in order to avoid possible confusion, that the presentation given here of this conjecture is quite different to that given in \cite{BSZ}, because the $esv$ ``rules'' defined in \cite{BSZ} (which contain the $esv$ map defined above) act on the whole asymptotic expansion of B-cycle graph functions, while here we will consider only the restriction to the first Laurent polynomial.
Note that the $esv$ conjecture would imply the part of Conjecture \ref{Conjsvmzv} concerning the coefficients $d_k^{(m,n)}(\Gamma)$ of the first Laurent polynomial, i.e. when $(m,n)=(0,0)$.

The \emph{esv} conjecture was checked for all graphs up to weight six, as well as for the ``crucial'' weight-seven example\footnote{This case could only be checked numerically, for about five-hundred digits.} (compare with eqn.~(\ref{Dweight7}))
\begin{align}
&\smallBcyc{G511}
 = 
 -\frac{31 T^7}{700539840 }
 -  \frac{ 5251 T^5 \zeta_2}{233513280}
 +  \frac{T^4 \zeta_3}{3888 }
 -  \frac{7405 T^3 \zeta_4}{598752}
  +  \frac{119 T^2 \zeta_5}{2592}
  +   \frac{31 T^2 \zeta_2 \zeta_3 }{864}  \notag \\
 & - \frac{ 11 T \zeta_3^2  }{ 216 } - \frac{15527 T \zeta_6}{10368}
+ \frac{21 \zeta_7}{32} +  \frac{67 \zeta_2 \zeta_5}{27}   + \frac{167 \zeta_3 \zeta_4}{48} 
 - \frac{ 23 \zeta_3 \zeta_5}{3 T}
 - \frac{80017 \zeta_8}{ 1296 T}
 + \frac{3 \zeta_2 \zeta_3^2}{T}  \notag \\
 &- \frac{25 \zeta_3^3}{4 T^2} + \frac{7115 \zeta_9}{144 T^2}
+ \frac{21 \zeta_2 \zeta_7}{T^2} + \frac{35 \zeta_4 \zeta_5}{6 T^2} - \frac{6613 \zeta_3 \zeta_6}{288 T^2}
 + \frac{75 \zeta_5^2}{4 T^3} - \frac{1245 \zeta_3 \zeta_7}{8 T^3}
  - \frac{48 \zeta_{3, 5} \zeta_2}{T^3}   \notag \\
  &+ \frac{ 443 \zeta_2 \zeta_3 \zeta_5}{T^3}
  - \frac{275 \zeta_3^2 \zeta_4}{8 T^3} + \frac{941869 \zeta_{10}}{ 5760 T^3}
   - \frac{ 9573 \zeta_{11}}{ 16T^4} - \frac{ 18 \zeta_{3,5,3} }{T^4} 
   - \frac{405 \zeta_3^2 \zeta_5}{  4T^4} 
   + \frac{195 \zeta_2 \zeta_3^3}{ 2T^4} \notag \\
    &+ \frac{27745 \zeta_5 \zeta_6}{ 48T^4}
   - \frac{ 3795 \zeta_4 \zeta_7}{ 16T^4}   + \frac{17731 \zeta_3 \zeta_8}{ 16T^4} + \frac{15875 \zeta_2 \zeta_9}{12 T^4}
  - \frac{2475 \zeta_5 \zeta_7}{4 T^5} - \frac{1125 \zeta_3 \zeta_9}{ 4 T^5} \notag \\
  &   + \frac{90 \zeta_{3, 5} \zeta_4}{T^5}      + \frac{450 \zeta_{3, 7} \zeta_2}{7 T^5}
      - \frac{ 165 \zeta_3 \zeta_4 \zeta_5}{2 T^5}  + \frac{3375 \zeta_2 \zeta_5^2}{7 T^5}
         + \frac{ 3335 \zeta_3^2 \zeta_6}{4 T^5}  + \frac{3960 \zeta_2 \zeta_3 \zeta_7}{7 T^5}  \notag \\
&    + \frac{ 93091945 \zeta_{12}}{11056 T^5}
- \frac{1575 \zeta_{13}}{T^6}   + \frac{13275 \zeta_2 \zeta_{11}}{4 T^6}  + \frac{7425 \zeta_4 \zeta_9}{8 T^6}
  + \frac{129465 \zeta_6 \zeta_7}{16 T^6}   \notag \\
  & + \frac{ 233525 \zeta_5 \zeta_8}{48 T^6} + \frac{160053 \zeta_3 \zeta_{10}}{64 T^6} 
+ \frac{15301285 \zeta_{14}}{768 T^7} \label{b511}
\end{align}
Let us now make two remarks in order to support the conjecture. First of all, it is straightforward to check that the appearence of the second Bernoulli polynomials in both eqn.~(\ref{FourierProp2}) and eqn.~(\ref{P=L+S}) implies that, for a fixed weight~$l$ graph, the coefficient of~$T^l$ in~(\ref{bLaurent}) is equal to the coefficient of~$(-2y)^l$ in~(\ref{dLaurent}). Moreover, note that the lowest power of~$T$ appearing in the expansion (\ref{bLaurent}) is~$-l$, while the lowest power of~$y$ appearing in the expansion~(\ref{dLaurent}) is~$1-l$. This means that the $esv$ conjecture implies that $sv(b_{-l}^{(0)}(\Gamma))=0$. We have seen in the proof of Theorem~\ref{Thm4} that there are two sources of contribution to the coefficient of~$T^{-l}$ in~(\ref{bLaurent}). The first one (obtained when~$s=0$) is given by $\prod \zeta(2)^{l_{i,j}}$, while the second one originates from the integrals of products (for all~$i,j$) of the kind
\begin{equation*}
L(z_{ij};\tau)^{l_{i,j}-1}S(z_{ij};\tau).
\end{equation*}
From the same kind of computation seen in the proof of Theorem \ref{Thm4} we can conclude that the contribution to the coefficient of~$T^{-l}$ (which is given by setting for all~$i,j$ $a_{i,j}=l_{i,j}-1-d_{i,j}$ in eqn.~(\ref{2ndContr})) is a rational linear combination of $\zeta(2(l-d))\zeta(2)^{d}$ for $0\leq d\leq l-1$. This implies that the coefficient of~$T^{-l}$ is a rational multiple of $\zeta(2l)$, which is indeed sent to~$0$ by the $sv$ map. In other words, we have proven:
\begin{prop}
For any weight~$l$ graph $\Gamma$ we have $esv(b_l^{(0)}(\Gamma)\,T^l)=d_l^{(0,0)}(\Gamma)\,y^l$ and $esv(b_{-l}^{(0)}(\Gamma)\,T^{-l})=0$.
\end{prop}

We want to conclude by explaining that the \emph{esv} conjecture should be related to Brown's recent construction of a new class of functions in the context of his research on mixed modular motives~\cite{BrownNewClassI}.
\begin{defn}[Brown 2017]
Let $f:\mathbb{H}\longrightarrow\mathbb{C}$ be a real analytic function. We call it \emph{modular of weights} $(r,s)$ if for every $\gamma=\left( \begin{array}{ccc}
a & b \\
c & d \end{array} \right)\in\mbox{SL}_2(\mathbb{Z})$ it satisfies
\begin{equation}\label{UN}
f(\gamma\tau)=(c\tau+d)^r(c\overline{\tau}+d)^sf(\tau).
\end{equation}
We denote $\mathcal{M}_{r,s}$ the space of modular functions of weights $(r,s)$ which admit an expansion of the form 
\begin{equation}\label{DOS}
f(q)\in\mathbb{C}[[q,\overline{q}]][y^{\pm 1}],
\end{equation}
where we recall that $y=\pi\,Im(\tau)$. We also define the bigraded algebra
\[
\mathcal{M}=\bigoplus_{r,s}\mathcal{M}_{r,s}.
\]
\end{defn}
Note that Theorem \ref{Thm3} implies that all modular graph functions belong to~$\mathcal{M}_{0,0}$.

The main result of \cite{BrownNewClassI} tells us, among other things, that there exists a subalgebra $\mathcal{M}\mathcal{I}^E\subset\mathcal{M}$ generated over $\mathcal{Z}^{\rm sv}$ by certain computable linear combinations of products of real and imaginary parts of iterated Eichler integrals of Eisenstein series, such that:
\begin{itemize}
\item It carries a grading by a certain \emph{M-degree} and a filtration by the length (number of integrations) of the iterated integrals, denoted by $\mathcal{M}\mathcal{I}^E_k\subset\mathcal{M}\mathcal{I}^E$.
\item Every element of $\mathcal{M}\mathcal{I}^E$ admits an expansion of the form
\[
f(q)\in\mathcal{Z}^{\rm sv}[[q,\overline{q}]][y^{\pm 1}]
\]
\item The sub-vector space of elements of fixed modular weights and M-degree $\leq m$ is finite dimensional.
\item Every element $F\in\mathcal{M}\mathcal{I}^E_k$ satisfies an inhomogeneous Laplace equation of the form
\[
(\Delta+w)F\in(\mathbb{E}+\overline{\mathbb{E}})[y]\times\mathcal{M}\mathcal{I}^E_{k-1}+\mathbb{E}\overline{\mathbb{E}}[y]\times \mathcal{M}\mathcal{I}^E_{k-2},
\]
where $\mathbb{E}$ denotes the space of holomorphic Eisenstein series for SL$_2(\mathbb{Z})$.
\end{itemize}

The elements of this algebra are called \emph{equivariant iterated Eichler integrals of Eisenstein series}. Their properties remind very much the conjectural properties of modular graph functions. In fact, in \cite{BrownNewClassI} it is conjectured that modular graph functions should belong to the weight $(0,0)$ subalgebra  of $\mathcal{M}\mathcal{I}^E$ (which would imply that Conjecture \ref{Conjsvmzv} is true). This is not surprising, if we think of the general phylosophy of string amplitudes: closed-string amplitudes are expected to be a single-valued non-holomorphic version of open-string amplitudes; since genus-one open-string amplitudes can be written in terms of (holomorphic) iterated Eichler integrals of Eisenstein series, we would therefore expect that the closed-string counterpart should be given by real-analytic single-valued (i.e. modular) analogues, and the most natural candidate is given by the elements of the algebra $\mathcal{M}\mathcal{I}^E$. However, the map which associates to a given iterated Eichler integral its equivariant image is rather involved beyond the simplest case of classical Eichler integrals\footnote{In this case, one simply takes the real part and lands on non-holomorphic Eisenstein series~\cite{MMV}.}, and this has initially discouraged the first attempts to use it to explicitly relate open-string integrals to closed-string integrals. On the other side, we have seen that by the surprisingly simple \emph{esv} conjecture we can obtain (the asymptotically big part of) modular graph functions from holomorphic graph functions, i.e. from special combinations of iterated Eichler integrals. We believe that, despite their apparently different origin, Brown's ``equivariant map'' and our \emph{esv} map should be related. Hopefully, understanding this link should pave the way towards proving both Conjecture \ref{Conjsvmzv} and the \emph{esv} conjecture. 

Finally, we want to remark that all this suggests, as anticipated in Section~\ref{sec:2}, that genus--$g$ superstring amplitudes for $g=0,1$ seem to be be related to mixed motives associated to moduli spaces of genus--$g$ Riemann surfaces. It would be extremely interesting to understand whether (and how) this will hold for higher genera, where the analogous mathematical structures have not been understood yet.

\vspace*{4mm}
\noindent
{\bf Acknowledgment.} We would like to thank KMPB for the organization of this successful conference. Moreover, we would like to thank C. Dupont, O. Schlotterer and M. Tapu\u{s}kovi\'c for useful comments on a first draft and J. Br\"odel and E. Garcia--Failde for their help with the figures. Our research was supported by a French public grant as part of the Investissement d'avenir project,
reference ANR-11-LABX-0056-LMH, LabEx LMH, and by the People Programme (Marie Curie Actions) of the European Union's Seventh Framework Programme
(FP7/2007-2013) under REA grant agreement n. PCOFUND-GA-2013-609102, through the PRESTIGE programme coordinated by Campus France.


\end{document}